\documentclass[aps,preprint]{revtex4}%
\usepackage{amsfonts}
\usepackage{amsmath}
\usepackage{amssymb}
\usepackage{graphicx}%
\begin{document}
\title{{\LARGE ON THE STABILITY OF OUR UNIVERSE}}
\author{Marcelo Samuel Berman$^{1}$ and \ Newton \ C. A. da Costa$^{1}$}
\affiliation{$^{1}$Instituto Albert Einstein/Latinamerica\ - Av. Candido Hartmann, 575 -
\ \# 17}
\affiliation{80730-440 - Curitiba - PR - Brazil }
\affiliation{emails: msberman@institutoalberteinstein.org , \ marsambe@yahoo.com, }
\affiliation{ncacosta@institutoalberteinstein.org , ncacosta@usp.br}
\keywords{Cosmology; Universe; Energy; Pseudotensor; Hawking; Mach.}\date{19 December 2010}

\begin{abstract}
\qquad\ \ \ \ \ \ \ \ \ \ \ \ \ \ We argue that the Robertson-Walker%
%TCIMACRO{\U{b4}}%
%BeginExpansion
\'{}%
%EndExpansion
s Universe is a zero-energy stable one, even though it may possess a
rotational state besides expansion.

\end{abstract}
\maketitle

\begin{center}

{\LARGE ON THE STABILITY OF OUR UNIVERSE}

\bigskip Marcelo Samuel Berman and Newton C. A. da Costa

\end{center}

\bigskip{\LARGE 1. Introduction}

\bigskip In his three best-sellers (Hawking, 1996; 2001; 2003), Hawking
describes inflation (Guth, 1981; 1998), as an accelerated expansion of the
Universe, immediately after the creation instant,while the Universe, as it
expands, borrows energy from the \ gravitational field to create more matter.
According to his \ description, the positive matter energy is exactly balanced
by the negative gravitational energy, so that the total energy is zero, and
that when \ the size of the Universe doubles, both the matter and
gravitational energies also double, keeping the total energy zero (twice
zero). Moreover, in the recent, next best-seller, Hawking and Mlodinow (2010)
comment that if it were not for the gravity interaction, one could not
\ validate a zero-energy Universe, and then, creation out of nothing would not
have happened.

In a previous paper Berman (2009c) has calculated the energy of the
Friedman-Robertson-Walker's Universe, by means of pseudo-tensors, and found a
zero-total energy. \textbf{Our main task will be to show that our possibly
rotating Robertson-Walker%
%TCIMACRO{\U{b4}}%
%BeginExpansion
\'{}%
%EndExpansion
s Universe is stable,in the sense that it has a reparametrized metric of
Minkowski's, while the latter has been shown to be the ground state of energy
level among possible universal metrics }(see Witten, 1981)\textbf{.}

\bigskip

\bigskip The zero-total-energy of the Roberston-Walker's Universe, and of any
Machian ones, have been shown by many authors. \bigskip It may be that the
Universe might have originated from a vacuum quantum fluctuation. By "vacuum",
we mean the spacetime of Minkowski. In support of this view, we shall show
that the pseudotensor theory (Adler et al, 1975) points out to a null-energy
for a rotating Robertson-Walker's Universe. Some prior work is mentioned,
(Tryon, 1973; Berman, 2006; 2006a; 2007; 2007a; 2007b; Rosen, 1994, 1995; York
Jr, 1980; Cooperstock, 1994; Cooperstock and Israelit, 1995; Garecki, 1995;
Johri et al.,1995; Feng and Duan,1996; Banerjee and Sen, 1997; Radinschi,
1999; Cooperstock and Faraoni, 2003). See also Katz (2006, 1985); Katz and Ori
(1990); and Katz et al (1997). Recent developments include torsion models (So
and Vargas, 2006), and, a paper by Xulu (2000).

The reason for the failure of non-Cartesian curvilinear coordinate energy
calculations through pseudotensors, resides in that curvilinear coordinates
carry non-null Christoffel symbols, even in Minkowski spacetime, thus
introducing inertial or fictitious fields that are interpreted falsely as
gravitational energy-carrying (false) fields.

\bigskip

{\LARGE 2. Reparametrization of Robertson-Walker's metric}

\bigskip

Consider first Robertson-Walker's metric, added by a temporal metric
coefficient which depends only on \ $t$\ \ . The line element (Gomide and
Uehara,1981), becomes:\ 

\bigskip

$ds^{2}=-\frac{R^{2}(t)}{\left(  1+kr^{2}/4\right)  ^{2}}\left[  d\sigma
^{2}\right]  +g_{00}\left(  t\right)  $ $dt^{2}$
\ \ \ \ \ \ \ \ \ \ \ \ \ \ \ . \ \ \ \ \ \ \ \ \ \ \ \ \ \ \ \ \ \ \ \ \ \ \ \ \ \ \ \ \ \ \ \ \ \ \ \ \ \ \ \ \ \ \ \ \ \ (1)

\bigskip

Of course, when \ $g_{00}=1$\ \ , the above equations reproduce conventional
Robertson-Walker's field equations.

\bigskip

We must mention that the idea behind Robertson-Walker's metric is the Gaussian
coordinate system. Though the condition \ \ $g_{00}=1$\ \ is usually adopted,
we must remember that, the resulting time-coordinate is meant as representing
proper time. If we want to use another coordinate time, we still keep the
Gaussian coordinate properties. Berman (2008a) has interpreted the generalized
metric as representing a rotating evolutionary model, with angular speed given by,

$\omega=\frac{\dot{g}_{00}}{2g_{00}}$

\bigskip

Consider the following reparametrization:

\bigskip

\bigskip$dx^{\prime2}\equiv\frac{R^{2}(t)}{\left(  1+kr^{2}/4\right)  ^{2}%
}dx^{2}$ \ \ \ \ \ \ \ \ \ \ \ \ \ \ \ \ \ \ \ \ \ \ \ \ \ \ \ \ , \ \ \ \ \ \ \ \ \ \ \ \ \ \ \ \ \ \ \ \ \ \ \ \ \ \ \ \ \ \ \ \ \ \ \ \ \ \ \ \ \ \ \ \ \ \ \ \ \ \ \ \ \ \ \ \ \ \ \ \ \ \ \ \ \ \ \ \ \ \ (2)

\bigskip$dy^{\prime2}\equiv\frac{R^{2}(t)}{\left(  1+kr^{2}/4\right)  ^{2}%
}dy^{2}$ \ \ \ \ \ \ \ \ \ \ \ \ \ \ \ \ \ \ \ \ \ \ \ \ \ \ \ \ , \ \ \ \ \ \ \ \ \ \ \ \ \ \ \ \ \ \ \ \ \ \ \ \ \ \ \ \ \ \ \ \ \ \ \ \ \ \ \ \ \ \ \ \ \ \ \ \ \ \ \ \ \ \ \ \ \ \ \ \ \ \ \ \ \ \ \ \ \ \ (3)

\bigskip$dz^{\prime2}\equiv\frac{R^{2}(t)}{\left(  1+kr^{2}/4\right)  ^{2}%
}dz^{2}$ \ \ \ \ \ \ \ \ \ \ \ \ \ \ \ \ \ \ \ \ \ \ \ \ \ \ \ \ , \ \ \ \ \ \ \ \ \ \ \ \ \ \ \ \ \ \ \ \ \ \ \ \ \ \ \ \ \ \ \ \ \ \ \ \ \ \ \ \ \ \ \ \ \ \ \ \ \ \ \ \ \ \ \ \ \ \ \ \ \ \ \ \ \ \ \ \ \ \ \ (4)

$dt^{\prime2}\equiv g_{00}(t)dt^{2}$
\ \ \ \ \ \ \ \ \ \ \ \ \ \ \ \ \ \ \ \ \ \ \ \ \ \ \ \ \ . \ \ \ \ \ \ \ \ \ \ \ \ \ \ \ \ \ \ \ \ \ \ \ \ \ \ \ \ \ \ \ \ \ \ \ \ \ \ \ \ \ \ \ \ \ \ \ \ \ \ \ \ \ \ \ \ \ \ \ \ \ \ \ \ \ \ \ \ \ \ \ \ (5)

\bigskip

\bigskip In the new coordinates, the generalized R.W.%
%TCIMACRO{\U{b4}}%
%BeginExpansion
\'{}%
%EndExpansion
s metric becomes:

\bigskip

$ds^{\prime2}=dt^{\prime2}-\left[  dx^{\prime2}+dy^{\prime2}+dz^{\prime
2}\right]  $ \ \ \ \ \ \ \ \ \ \ \ \ \ \ \ \ \ \ \ \ \ \ \ \ \ \ \ \ . \ \ \ \ \ \ \ \ \ \ \ \ \ \ \ \ \ \ \ \ \ \ \ \ \ \ \ \ \ \ \ \ \ \ \ \ \ \ \ \ \ \ \ \ \ \ \ \ (6)

\bigskip

This is Minkowski's metric.

\bigskip

\bigskip

{\LARGE 3. Energy and stability of the Robertson-Walker's metric}

\bigskip

\bigskip Even in popular Science accounts (Hawking, 1996; 2001; 2003\bigskip;
--- and Moldinow, 2010; Guth,1998), it has been generally accepted that the
Universe has zero-total energy. The first such claim, seems to be due to
Feynman (1962-3). Lately, Berman (2006, 2006 a) has proved this result by
means of simple arguments involving Robertson-Walker's metric for any value of
the tri-curvature ( $0,-1,1$ ).

\bigskip

Berman and Gomide (2010,2011) has recently shown that the generalized
Robertson-Walker's metric yielded a zero-energy pseudotensorial result. The
same authors showed that the result applied in case of a rotating and
expanding Universe.

\bigskip The equivalence principle, says that at any location, spacetime is
\ (locally) flat, and a geodesic coordinate system may be constructed,
\textit{where the Christoffel symbols are null. The pseudotensors are, then,
at each point, null. But now remember that our old Cosmology requires a
co-moving observer at each point}. \textit{It is this co-motion that is
associated with the geodesic system, and, as RW%
%TCIMACRO{\U{b4}}%
%BeginExpansion
\'{}%
%EndExpansion
s metric is homogeneous and isotropic, for the co-moving observer, the
zero-total energy density result, is repeated from point to point, all over
spacetime. Cartesian coordinates are needed, too, because curvilinear
coordinates are associated with fictitious or inertial forces, which would
introduce inexistent accelerations that can be mistaken \ additional
gravitational fields (i.e., that add to the real energy). Choosing Cartesian
coordinates is not analogous to the use of center of mass \ frame in Newtonian
theory, but the null results for the spatial components of the
pseudo-quadrimomentum show compatibility. }

\bigskip\bigskip

\bigskip Witten (1981) proved that within a semiclassical approach,
Minkowski's space was in the ground state of energy, which was zero-valued. He
also showed that in Classical General Relativity, this space also was the
unique space of lowest energy. This last result was obtained with spinor
calculus, and thus could be extended to higher dimensions whenever spinors
existed. The proof was obtained through the study of the limit
\ \ \ $h\rightarrow0$\ \ \ \ of a supergravity argument by Deser and
Teitelboim (1977) and by Grisaru (1978), where \ \ \ $h$\ \ \ stands for
Planck's constant.

\bigskip

The conclusion of Witten was that Minkowski's space was also stable, because
perturbations in the form of gravitational waves should not decrease the total
energy, because it is known that gravitational waves have positive energy. We
now conclude that our Universe is also stable, due to the reparametrization
above. But, first, let us deal with some conceptual issues.

\bigskip

\bigskip We have three kinds of stability criteria: 1) . Since a physical
system shows a tendency to decay into its state of minimum energy, the
criterion states that \ the system should not be able to collapse into a
series of infinitely many possible negative levels of energy. There should be
a minimum level, usually zero-valued, which is possible for the physical
system.; 2) . The matter inside the system must not be possibly created out of
nothing,or else, the bodies should have positive energy.; 3) . "Small"
disturbances should not alter a state of equilibrium of the system (it tends
to return to the original equilibrium state). In the case of the Universe,
disturbances, of course, cannot be external.

According with our discussion, the rotating Robertson-Walker%
%TCIMACRO{\U{b4}}%
%BeginExpansion
\'{}%
%EndExpansion
s Universe is locally and globally stable, whenever Classical Physics is
concerned. Now, Berman and Trevisan (2010), have shown that Classical General
Relativity can be used to describe the scale-factor of the Universe even
inside Planck%
%TCIMACRO{\U{b4}}%
%BeginExpansion
\'{}%
%EndExpansion
s zone, provided that we consider that the calculated scale-factor behaviour
reflects an average of otherwise uncertain values, due to Quantum fluctuations.

\bigskip

\bigskip{\LARGE 4. Final Comments and Conclusions}

\bigskip

\bigskip Berman and Gomide (2010,2011) have obtained a zero-total energy proof
for a rotating expanding Universe. The zero result for the spatial components
of the energy-momentum-pseudotensor calculation, are equivalent to the choice
of a center of Mass reference system in Newtonian theory, likewise the use of
comoving observers in Cosmology. It is with this idea in mind, that we are led
to the energy calculation, yielding zero total energy, for the Universe, as an
acceptable result: we are assured that we chose the correct reference system;
this is a response to the criticism made by some scientists which argue that
pseudotensor calculations depend on the reference system, and thus, those
calculations are devoid of physical meaning.

\bigskip

Related conclusions should be consulted (see all Berman's references and
references therein). As a bonus, we can assure that there was not an initial
infinite energy density singularity, because attached to the zero-total energy
conjecture, there is a zero-total energy-density result, as was pointed by
Berman elsewhere (Berman, 2008). The so-called total energy density of the
Universe, which appears in some textbooks, corresponds only to the
non-gravitational portion, and the zero-total energy density results when we
subtract from the former, the opposite potential energy density.

\bigskip

As Berman (2009d; f) shows, we may say that the Universe is
\ \emph{singularity -free }, and was created \emph{ab-nihilo },;in particular,
there is no zero-time infinite energy-density singularity.

\bigskip

\bigskip Rotation of the Universe and zero-total energy were verified for
Sciama's linear theory, which has been expanded, through the analysis of
radiating processes, by one of the present authors (Berman, 2008d; 2009e).
There, we found Larmor's power formula, in the gravitational version,that
leads to the correct constant power relation for the Machian Universe.
However, we must remember that in local Physics, General Relativity deals with
quadrupole radiation, while Larmor is a dipole formula; for the Machian
Universe the resultant constant power is basically the same, either for our
Machian analysis or for the Larmor and general relativistic formulae.

\bigskip

Referring to rotation, it could be argued that cosmic microwave background
radiation should show evidence of quadrupole asymmetry and it does not, but
one could argue that the angular speed of the present Universe is too small to
be detected; also, we must remark that CMBR deals with null geodesics, while
Pioneers' anomaly, for instance, deals with time-like geodesics. In favor of
evidence on rotation, we remark neutrinos' spin, parity violations, the
asymmetry between matter and anti-matter, left-handed DNA-helices, the fact
that humans and animals alike have not symmetric bodies, the same happening to molluscs.

\bigskip

We predict that chaotic phenomena and fractals, rotations in galaxies and
clusters, may provide clues on possible left handed preference through the Universe.

\bigskip

Berman and Trevisan (2010) have remarked that creation out-of-nothing seems to
be supported by the zero-total energy calculations. Rotation was included in
the derivation of the zero result by Berman and Gomide (2010). We could think
that the Universes are created in pairs, the first one (ours), has negative
spin and positive matter; the second member of the pair, would have negative
matter and positive spin: for the ensemble of the two Universes, the total
mass would always be zero; the total spin, too. The total energy (twice zeros)
is also zero.

{\LARGE \bigskip}

\bigskip Hawking and Mlodinow (2010) conclude their book with a remark on the
fact that the Universe is locally stable, but globally unstable because
spontaneous creation is the reason why the Universe exists, and new creations
like this may still happen. Of course, this is a question of interpretation.

We now want to make a conjecture related to the stability criteria of last Section.

\bigskip A physical system is not "chaotic", if small perturbations in its
initial state do not originate "large" variations in its future behaviour.
According to our discussion, the Robertson-Walker%
%TCIMACRO{\U{b4}}%
%BeginExpansion
\'{}%
%EndExpansion
s Universe, with or without rotation, is locally and globally stable under the
three criteria. As its total energy is zero, we conjecture that this type of
Universe is not globally chaotic, and that the three criteria for stability
imply that any such system cannot be globally chaotic altogether. We remark
nevertheless, that because Einstein%
%TCIMACRO{\U{b4}}%
%BeginExpansion
\'{}%
%EndExpansion
s field equations are non-linear,chaos is not forbidden in a local sense.

We regret that the name of a basic result in General Relativity Theory, is
called "positive energy theorem " instead of the "non-negative energy theorem".

\bigskip

.

{\LARGE Acknowledgements}

\bigskip

The authors thank Marcelo Fermann Guimar\~{a}es, Nelson Suga, Mauro Tonasse,
Antonio F. da F. Teixeira, and for the encouragement by Albert, Paula, Geni
(MSB) and Neusa (NC).

\bigskip

\bigskip

\bigskip{\Large References}

\bigskip

Adler, R.J.; Bazin, M.; Schiffer, M. (1975) - \ \textit{Introduction to
General Relativity, }2$^{nd}$ Edition, McGraw-Hill, New York.

Banerjee, N.; Sen, S. (1997) - Pramana J.Phys., \textbf{49}, 609.

Berman, M.S. (1981, unpublished) - M.Sc. thesis, Instituto Tecnol\'{o}gico de
Aeron\'{a}utica, S\~{a}o Jos\'{e} dos Campos, Brazil.Available online, through
the federal government site \ www.sophia.bibl.ita.br/biblioteca/index.html
(supply author%
%TCIMACRO{\U{b4}}%
%BeginExpansion
\'{}%
%EndExpansion
s surname and keyword may be "pseudotensor"or "Einstein").

\bigskip Berman, M. S. (2006) - \textit{Energy of Black-Holes and Hawking%
%TCIMACRO{\U{b4}}%
%BeginExpansion
\'{}%
%EndExpansion
s Universe}, in Chapter 5 of \textit{Trends in Black Hole Research}, ed by
Paul V. Kreitler, Nova Science, New York.

Berman, M. S. (2006a) - \textit{Energy, Brief History of Black-Holes,and
Hawking%
%TCIMACRO{\U{b4}}%
%BeginExpansion
\'{}%
%EndExpansion
s Universe}, in Chapter 5 of: \textit{New Developments in Black Hole
Research}, ed by Paul V. Kreitler, Nova Science, New York.

\bigskip Berman, M.S. (2007) - \textit{Introduction to General Relativity and
the Cosmological Constant Problem, }Nova Science, New York.

Berman, M.S. (2007a) - \textit{Introduction to General Relativistic and Scalar
Tensor}

\textit{Cosmologies }, Nova Science, New York.

Berman, M.S. (2007b) - \textit{The Pioneer Anomaly and a Machian Universe} -
Astrophysics and Space Science, \textbf{312}, 275. Los Alamos Archives, http://arxiv.org/abs/physics/0606117.

Berman, M.S. (2008a) - \textit{A General Relativistic Rotating Evolutionary
Universe, }Astrophysics and Space Science, \textbf{314, }319-321.

\bigskip Berman, M.S. (2008c) - \textit{A Primer in Black Holes, Mach's
Principle and Gravitational Energy, }Nova Science, New York.

Berman, M.S. (2008d) - \textit{On the Machian Origin of Inertia}, Astrophysics
Space Science, \textbf{318}, 269-272. Los Alamos Archives, http://arxiv.org/abs/physics/0609026

\bigskip Berman, M.S. (2009) - \textit{General Relativistic Singularity-free
Cosmological Model, }Astrophysics and Space Science, \textbf{321}, 157-160.

\bigskip

Berman, M.S. (2009c) - \textit{On the zero-energy Universe, }International
Journal of Theoretical Physics, \textbf{48}, 3278.

Berman, M.S. (2009d) - \textit{Why the initial infinite singularity of the
Universe is not there?, }International Journal of Theoretical Physics,
\textbf{48}, 2253.

Berman, M.S. (2009e) - \textit{On Sciama's Machian Cosmology, }International
Journal of Theoretical Physics, \textbf{48}, 3257.

Berman, M.S. (2009f) - \textit{General Relativistic Singularity-Free
Cosmological Model, }Astrophysics and Space Science, \textbf{321}, 157. Los
Alamos Archives http://arxiv.org/abs/0904.3141.

Berman, M.S.; Gomide, F.M. (2010) - \textit{General Relativistic Treatment of
Pioneers anomaly, } submitted. See Los Alamos Archives \ http://arxiv.org/abs/1011.4627

\bigskip Berman, M.S.; Gomide, F.M. (2011)-\textit{On the rotation of the
Zero-energy Expanding Universe,}in \textit{The Big-Bang-Theory and
Assumptions,}ed. by F.Columbus,Nova Science Publishers, New York, to be published.

\begin{description}
\item Berman, M.S.; Trevisan, L.A. (2010) - International Journal of Modern
Physics, \textbf{D19, }1309-1313.  Los Alamos Archives http://arxiv.org/abs/gr-qc/0104060
\end{description}

Cooperstock, F.I. (1994) - GRG \textbf{26}, 323.

Cooperstock, F.I.; Israelit, M. (1995) - \textit{Foundations of Physics},
\textbf{25}, 631.

\bigskip Cooperstock, F.I.; Faraoni, V.(2003) - Ap.J. 587,483.

Deser, S.; Teitelboim, C. (1977) - Phys. Rev. Lett. \textbf{39}, 249.

\begin{description}
\item Feng, S.; Duan, Y. (1996) - Chin. Phys. Letters, \textbf{13}, 409.
\end{description}

Feynman, R. P. (1962-3) - \textit{Lectures on Gravitation} , Addison-Wesley, Reading.

Garecki, J.\ (1995) - GRG, \textbf{27}, 55.

Gomide, F.M.; Uehara, M. (1981) - Astronomy and Astrophysics, \textbf{95}, 362.

Grisaru, M. (1978) - Phys. Lett. \textbf{73B}, 207.

Guth, A. (1981) - Phys. Rev. \textbf{D23}, 347 .

Guth, A. (1998) - \textit{The Inflationary Universe,}\ Vintage, New York, page 12.

\bigskip Hawking, S. (1996) - \textit{The Illustrated A Brief History of Time,
}Bantam Books, New York, pages 166-167.

Hawking, S. (2001) - \textit{The Universe in a Nutshell, }Bantam Books, New
York, pages 90-91.

Hawking, S. (2003) - \textit{The Illustrated Theory of Everything, }Phoenix
Books, Beverly Hills, page 74.

Hawking, S.; Mlodinow, L. (2010) - \textit{The Grand Design, }Bantam Books,
New York.

Johri, V.B.; et al. (1995) - GRG, \textbf{27}, 313.

Katz, J. (1985) - Classical and Quantum Gravity \textbf{2}, 423.

Katz, J. (2006) - Private communication.

Katz, J.; Ori, A. (1990) - Classical and Quantum Gravity\textbf{ 7}, 787.

Katz, J.; Bicak, J.; Lynden-Bell, D. (1997) - Physical Review \textbf{D55}, 5957.

Radinschi, I. (1999) - Acta Phys. Slov., \textbf{49}, 789. Los Alamos
Archives, gr-qc/0008034.

Rosen, N.(1994) - \textit{Gen. Rel. and Grav.} \textbf{26}, 319.

Rosen, N.(1995) - GRG, \textbf{27}, 313.

Sciama, D.W. (1953) - M.N.R.A.S., \textbf{113}, 34.

So, L.L.; Vargas, T. (2006) - Los Alamos Archives, gr-qc/0611012 .

Tryon, E.P.(1973) - Nature, \textbf{246}, 396.

\begin{description}
\item Witten, E. (1981) - \textit{A New Proof of the Positive Energy Theorem,
}Commun. Math. Phys. \textbf{80}, 381-402.
\end{description}

\bigskip Xulu, S. (2000) - International Jounal of Theoretical Physics,
\textbf{39}, 1153. Los Alamos Archives, gr-qc/9910015 .

York Jr, J.W. (1980) - \textit{Energy and Momentum of the Gravitational
Field}, in \textit{A Festschrift for Abraham Taub}, ed. by F.J. Tipler,
Academic Press, N.Y.\bigskip

\end{document}